\documentclass[aps,prl,twocolumn,showpacs,superscriptaddress]{revtex4}
\usepackage{graphicx,amssymb,amsfonts,amsmath}

\begin{document}
\newcommand{\AR}[1]{{\bf AR} {\it #1} {\bf END}}
\newcommand{\OLD}[1]{{\bf OLD} {\tiny #1}}

\title{Current induced decoherence in the multichannel Kondo problem} \author{Aditi Mitra}
\affiliation{Department of Physics, New York University, 4 Washington
  Place, New York, New York 10003, USA} \author{Achim Rosch}
\affiliation{Institute for Theoretical Physics, University of Cologne,
  50937 Cologne, Germany} \date{\today}


\begin{abstract}
The properties of a local spin $S=1/2$ coupled to
 $K$ independent wires  is studied in
the presence of bias voltages which drive the system out of thermal
equilibrium. For $K\gg 1$, a perturbative renormalization group
approach is employed to construct the voltage dependent scaling
function for the conductance and the T-matrix.  In contrast to
the single-channel case, the Kondo resonance is split even by bias
voltages small compared to the Kondo temperature $T_K$, $V \ll T_K$.
Besides the applied voltage $V$, the current induced
decoherence rate $\Gamma \ll V$ controls the physical properties
of the system. While the presence of $V$ changes the structure of
the renormalization group considerably, decoherence turns out
to be be very effective in prohibiting the flow towards new
nonequilibrium fixed points even in variants of the Kondo model
where currents are partially suppressed.
\end{abstract}

\pacs{71.10.-w,71.27.+a,73.63.-b,05.70.Ln}

\maketitle

The problem of a local spin $S$ exchange coupled to $K$ independent
conduction electron reservoirs, inspite of its apparent simplicity
shows a rich variety of phases. For example when $K$=$2S$, the local
spin is completely screened and the system behaves as a
Fermi-liquid~\cite{Hewson}.  In contrast for $K>2S$, the spin is
over-screened, and the system exhibits non-Fermi liquid
behavior characterized by a zero temperature entropy
$S=\ln{g}$ where $g$ is a non-integer~\cite{Nozieres80,Andrei84,Affleck93,Gan94}.
Recently
the over-screened Kondo problem for $K$=$2,S$=$1/2$ was realized in a controlled
experimental set-up~\cite{Potok07}, opening up the possibility of
probing these exotic systems in the far out of equilibrium
regime.

An important question in the study of any strongly correlated system
is the possibility of realizing new fixed points and scaling
behavior by driving the system out of equilibrium. In~\cite{2CKneq} it was proposed that
the single channel Kondo model in the large bias limit should flow to a new
fixed point which is characterized by a change in the number of independent screening
channels as coherent scattering processes between leads are
prohibited by the bias voltage.
However subsequent studies~\cite{Glazman00,Rosch01,Flow,Mitra07,RealtimeRG}
of the nonequilibrium single-channel Kondo model ruled out such a nonequilibrium regime due
to current induced decoherence which qualitatively plays the role of an effective temperature
(or an infra-red cutoff),
and can be as large as the voltage in the single-channel Kondo model.

In this paper we study the over-screened Kondo problem when a local
spin $S$=$1/2$ is coupled to $K$ independent current carrying channels.
In the limit $K \gg 1$, the intermediate coupling non-Fermi liquid
fixed point can be studied within renormalized Keldysh perturbation theory in
$1/K$.  We show that the current induced decoherence in this model
is ${\cal O}(1/K)$ and thus considerably suppressed. In spite of this we find that
decoherence is highly effective in cutting off the flow to any nonequilibrium
fixed points. We present results for the voltage dependence of the
conductance and the T-matrix, the latter being related to the
local density of states that can be probed in a
tunneling experiment.

The Hamiltonian is $H=H_0 + V_X$ where
\begin{eqnarray}
&&H_0 = \sum_{\overset{k\sigma \alpha=L,R}{m=1\ldots K}}
(\epsilon_k -\mu_{\alpha}^m)c_{k\sigma m \alpha}^{\dagger} c_{k\sigma m\alpha}
\end{eqnarray}
represents $K$ independent free electron reservoirs labeled by
$m$. Each of the $K$ reservoirs
is split into a left ($L$) and a right ($R$) part which can be
maintained at different chemical potentials $\mu_L^m\neq \mu_R^m$.
 The local spin is
coupled to the spin density of each wire via the exchange interaction,
\begin{eqnarray}
&&V_X =  \frac{1}{2} \vec{S}\cdot
\sum_{\overset{\alpha,\beta=L,R;k,k^{\prime}}{\sigma\sigma^{\prime};m=1\ldots K}}
J_{\alpha \beta}
c^{\dagger}_{k\sigma m \alpha}\vec{\sigma}_{\sigma \sigma^{\prime}}c_{k^{\prime}\sigma^{\prime}m\beta}
\label{H}
\end{eqnarray}
Here, $J_{LR}$ connects the $L$ and
$R$ leads so that a net current can flow within each reservoir when
$\mu_{L}^m\neq \mu_{R}^m$. However there is no flow of current from channel $m$ to channel $m'\neq m$.

We evaluate two physical quantities, the current through the $m$-th lead
$\hat{I}_m $=$e \frac{dN_{Lm}}{dt}$
which is given by
$\hat{I}_m
$=$ -i\frac{e}{\hbar}J_{LR}\frac{\vec{S}}{2}\cdot \sum_{kk^{\prime}\sigma\sigma^{\prime}}
\left(c^{\dagger}_{k\sigma L m} \vec{\sigma}_{\sigma \sigma^{\prime}}
c_{k^{\prime}\sigma^{\prime}Rm} - h.c.\right)$,
and the $T$-matrix of the electrons in the $m$-th lead defined by
$G^{mkk^{\prime}\alpha \alpha^{\prime}}_R
$=$\delta_{kk^{\prime}}\delta_{\alpha \alpha^{\prime}}
G^{0mk}_R + G^{0mk\alpha}_R T_R^{m\alpha \alpha^{\prime}}
G^{0mk^{\prime}\alpha^{\prime}}_R$,
where
$G^{m kk^{\prime}\alpha \alpha^{\prime}}_R$ is the retarded propagator for the electrons in
the $m$-th lead and the T-matrix is the expectation value of the
composite operator $T^{m\alpha \alpha^{\prime}}_{a b}(t,t^{\prime})$=$ -i T_C\langle O_a(t)
O^{\dagger}_b(t^{\prime})\rangle$ where
$
O_a(t)$=$\frac{1}{2}\sum_{\alpha_1 k_1\sigma_1}
J_{\alpha \alpha_1} \vec{S}(t)\cdot \vec{\sigma}_{\sigma \sigma_1}
c_{m\alpha_1 k_1 \sigma_1 a }(t)
$, $a,b$=$\pm$ denote Keldysh labeling and $T_C$ denotes Keldysh time-ordering~\cite{Keldysh}.
Formally, under assumptions that the interactions $V_X$ was switched on
at $t$=$-\infty$, the expectation value of any operator $\hat{O}$ at
time $t$=$0$ is~\cite{Keldysh}
$\langle \hat{O}(0)\rangle$=$
Tr\left[\left(\tilde{T} e^{i\int_{-\infty}^0 dt^{\prime}
V_X(t^{\prime})}\right)\hat{O}\left({T} e^{-i\int_{-\infty}^0 dt
V_X(t)}\right) {\rho}\right]
$
where ${\rho}$ is the initial density matrix at $t$=$-\infty$ and $V_X(t)$=$ e^{i H_0 t}
V_X e^{-i H_0 t}$. We assume
$\rho $=$ \rho_{leads} \otimes \rho_S$
where $\rho_S$ is the density matrix for the free
spin, and $\rho_{leads}$ is such that
$\langle c^{\dagger}_{k\sigma m \alpha} c_{k\sigma m\alpha} \rho_{leads}\rangle $=$
f(\epsilon_k -\mu_{\alpha}^m)$,
$f$ being the Fermi function. We also assume that the leads have a uniform density of states
$\nu$ and a bandwidth $2D$.

{\bf Identical voltage drops across the K-wires:}
Let us suppose that all the $K$ wires have the same voltage drop
$V$=$\mid \mu_L- \mu_R \mid$
applied across them. Then a perturbative treatment to two-loop order, where only the diagrams which
are ${\cal O}(K)$ at two loop are kept gives
the following result for
the conductance $G_m = \partial I_m/\partial V$,
\begin{eqnarray}
&&G_m = \frac{3\pi}{4}\frac{e^2}{\hbar}\left(\nu J_{LR}\right)^2\left[1 + 2 \nu\left(J_{LL}
+ J_{RR}\right) \ln{\frac{D}{V}} \right. \nonumber \\
&&\left. - K \ln\frac{D}{V} \sum_{\alpha\beta=L,R}\theta(D-\mid
  \mu_\alpha -
\mu_\beta \mid)
(\nu J_{\alpha\beta})^2 \right] \label{G4m}
\end{eqnarray}
In addition the imaginary part of the $T$-matrix of the $m$-th wire is found to be,
\begin{eqnarray}
&&-\pi\nu {\rm Im}\left[T^{m\alpha\alpha^{\prime}\sigma}_{R}(\Omega)\right]=
\frac{3 \pi^2}{16}\nu^2 \sum_{\alpha_1}
\left[J_{\alpha \alpha_1} J_{\alpha_1\alpha^{\prime}} \right. \nonumber\\
&&\left. + 2\nu\sum_{\alpha_2}
J_{\alpha \alpha_1} J_{\alpha_1\alpha_2}J_{\alpha_2\alpha^{\prime}} \ln\left(\frac{D}{\mid \Omega
-\mu_{\alpha_1}\mid}\right) \right. \nonumber\\
&&\left.
-K  J_{\alpha \alpha_1} J_{\alpha_1\alpha^{\prime}}
\ln\left(\frac{D}{\mid \Omega - \mu_{\alpha_1}\mid}\right) \right. \nonumber \\
&&\left. \times \sum_{\gamma \delta} \nu^2 J_{\gamma\delta}^2
\theta(D-\mid \mu_{\gamma}-\mu_{\delta}\mid)
\right] \label{Tpert3}
\end{eqnarray}

According to Eq.~(\ref{G4m})
all logarithmic singularities in the conductance are cutoff by
the voltage. This reflects the fact that
in a model where the cutoff $D$ is much smaller than the voltage
difference $V$ no current will flow by energy conservation. In
contrast, the calculation of the $T$-matrix shows that even in this
regime resonant spin-flip processes lead to logarithmic
renormalizations. For the
renormalization group (RG)
analysis it will be important, that some $J_{LR}$ processes do not
contribute for $D<V$ as described by the $\theta$ function terms in Eq.~(\ref{Tpert3}).

We will formulate the
RG equations in terms of dimensionless couplings,
$g_{ab}$=$\nu J_{ab}$ assuming symmetric couplings, $g_{LL}$=$g_{RR}
$=$g_d$ but we will allow for $g_{LR} \neq g_d$. While for a simple
 Anderson model one always obtains $g_{LR}$=$g_d$, this
is not valid for more complex models.

When deriving two-loop equations in full generality for a
non-equilibrium situation using, e.g., functional renormalization
group~\cite{FRG}, flow equations~\cite{Flow} or real-time
renormalization group approaches~\cite{RealtimeRG}, it is necessary to take
into account the full energy dependence of interaction
vertices. Even at the one-loop level it is useful~\cite{Freqdep} to keep track of
how the coupling constants $g$ depend on the energy of the incoming
electron. To leading order in $1/K$, however, and for the quantities considered
in this paper, we believe that it is sufficient to use a simpler
version guided by results from perturbation theory, concentrating on a few on-shell coupling
constants. Already in perturbation theory to order $J^3$ the
structure of logarithmic corrections depends on which physical quantity
is considered. The simplest case is the conductance where all
logarithmic corrections up to the order considered are cutoff by the
voltage. Thus for the conductance it is sufficient to use the well-known equilibrium
RG equations~\cite{Hewson}
\begin{eqnarray}
\frac{dg_d}{d\ln D}& =& -\left[\left(g_d^2 + g_{LR}^2\right)
-\frac{K}{2}g_d\left(2 g_d^2 + 2 g_{LR}^2\right)\right]\nonumber \\
\frac{dg_{LR}}{d\ln D} &=& -\left[2 g_d g_{LR}
-\frac{K}{2}g_{LR}\left(2 g_d^2 + 2 g_{LR}^2\right)
\right]
\end{eqnarray}
supplemented by the condition that the RG flow is cutoff
at $D$=$V$.

However, when one requires instead that the
$T$-matrix $T^{\alpha \alpha}(\mu_\alpha)$, evaluated at the
Fermi energy of the lead, remains invariant under RG,
one obtains the RG equations
\begin{eqnarray}
\frac{dg_d}{d\ln D}\! &=& \! -\left[\left(g_d^2 + g_{LR}^2 \Theta_V \right)
-\frac{K}{2}g_d\left(2 g_d^2 + 2 g_{LR}^2  \Theta_V \right)\right]\nonumber\\
\frac{dg_{LR}}{d\ln D}\! &=& \! -\left[g_d g_{LR}(1+ \Theta_V)
-\frac{K}{2}g_{LR}\left(2 g_d^2 + 2 g_{LR}^2 \Theta_V \right)
\right]\nonumber\\
\label{rgfull}
\end{eqnarray}
Here the factors $\Theta_V=\Theta(D-V)$ take into account that part of
the logarithms and therefore the
RG-flow is cutoff by the voltage when the cutoff becomes smaller than
the voltage. Therefore the equilibrium RG flow is modified at the
scale $V$ in a way which can be simply understood as a suppression of
all resonant scattering processes from one to the other chemical
potential by $g_{LR}$.

Note that Eq.~(\ref{rgfull}) is not valid in the limit $D \to 0$
because the RG flow is ultimately cutoff by the decoherence processes
as had been pointed out previously~\cite{Glazman00,Rosch01,Flow,Mitra07}. A careful analysis of the
decoherence rates (including 3 loop diagrams) has been performed by
Schoeller and Reininghaus in~\cite{RealtimeRG}
by investigating the time evolution of the density
matrix of the impurity. Their analysis shows that within the precision
of a two-loop calculation (i.e. within our model to leading order in
$1/K$), the decoherence rate is given by the simple formula used
previously~\cite{Glazman00,Rosch01} for one-loop calculations. Up to prefactors of order
$1$, irrelevant for our discussion, the RG flow is ultimately cutoff
at the scale $\Gamma$,
\begin{eqnarray}
\Gamma = \pi K g_{LR}(V)^2 V \label{gamdef}
\end{eqnarray}
where $g_{LR}(V)$ is the running coupling constant at the scale $D$=$V$.
The factor $K$ arises as each of the $K$ channels contributes to
the dephasing. $\Gamma$ is the Korringa spin-relaxation rate~\cite{Glazman00,Rosch01,Flow,Mitra07}
and is proportional to the current.

Let us make the assumption $g_d$=$g_{LR}$ (we will relax this condition later).
For the regime $D>V$ it is convenient to define $g$=$2 g_d$
which obeys the RG equation
\begin{eqnarray}
\frac{dg}{d\ln D} = -\left[g^2 - \frac{K}{2}g^3\right] =\beta(g)\label{rgeq1}
\end{eqnarray}
The above equation has the well known
fixed point at $g^*$=$2/K$~\cite{Nozieres80}, whereas the
scaling dimension of the leading irrelevant operator is
$\Delta$=$\beta^{\prime}(g^*)$=$\frac{2}{K}$ valid for $K\gg 1$
(the exact result \cite{Affleck93} is $\Delta=2/(K+2)$).
Integrating Eq.~(\ref{rgeq1}) upto an energy-scale $D$,
and defining the Kondo temperature as
$
T_{K}$=$D_0 \left(\frac{g_0}{g^*}\right)^{K/2} e^{-1/g_0} $, ($D_0,g_0$ being the initial bandwidth
and coupling constant respectively), one gets~\cite{Gan94}
\begin{eqnarray}
&&|g(D) - g^* | \nonumber \\
&&= | g_0-g^{*}| \left[\frac{g(D)}{g^*}\right]^{K\Delta/2}
\left(\frac{D}{{T}_K}\right)^{\Delta}
e^{-\Delta/g(D)} \label{rgeq3a}
\end{eqnarray}
\begin{figure}
\includegraphics[totalheight=5cm]{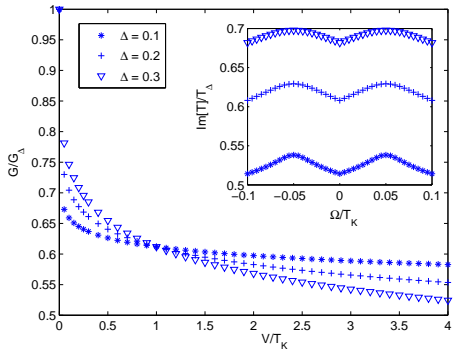}
\caption{(Color online) Main panel:
The conductance in the scaling limit and for several different $\Delta$=$2/K$.
$G_{\Delta}$=$\left(\frac{2e^2}{h}\right)\frac{3\pi^2}{16}\Delta^2$. Inset: $Im[T(\Omega)]$
for $V/T_K$=$0.1$ and the same $\Delta$ as in the main panel. $T_{\Delta}$=$3\pi^2\Delta^2/32$.}
\label{Gcr}
\end{figure}

Eq.~(\ref{rgeq3a}) is valid for arbitrary $D/{T}_K$.
For $D \gg {T}_K$, i.e. for $g \ll g^*$,
we obtain
$g(D) \simeq \frac{1}{\ln D/{T}_K} +\frac{K}{2}
\frac{\ln(2/K \ln D/T_K)}{(\ln D/T_K)^2} + {\cal O}(\frac{1}{\ln D/T_K})^3$.
For $D \ll {T}_K $, setting $g(D)$=$g^*$
on the r.h.s of Eq.~(\ref{rgeq3a}) one gets
\begin{eqnarray}
g(D) = g^* \left(1-\xi
  \left(\frac{D}{{T}_K}\right)^{\Delta}\right)
 \label{gRsol}
\end{eqnarray}where $\xi = \left(1-g_0/g^* \right)
e^{-\Delta/g^*}$. In the scaling limit  $g_0\rightarrow 0$, and for
large $K$, $\xi$ takes the universal value
$\xi = 1/e$.

Integrating upto the energy scale $D$=$V$, the logarithms in the
conductance are resummed  giving $G_m \approx \frac{3\pi}{4}\frac{e^2}{\hbar} g_{LR}^2(V)$.
In particular the result for $V\ll T_K$ is
\begin{eqnarray}
G_m \approx \frac{3\pi}{4} \frac{e^2}{\hbar} \frac{g^{*2}}{4}\left[1 - 2\xi
\left(\frac{V}{{T}_K}\right)^{\Delta}\right] \label{conductance}
\end{eqnarray}
where we have dropped higher order terms $\sim \left(\frac{V}{{T}_K}\right)^{2\Delta}$.
Note that the conductance near the fixed point is a quantity of ${\cal O}(1/K^2)$.
For arbitrary $V/T_K$ and in the scaling limit, the
conductance is given by the universal function
$G/G_{\Delta}=1/\left[1+W\left(e^{-1}(V/T_K)^{\Delta}\right)\right]^2$ where
$G_{\Delta}$=$
\left(\frac{2e^2}{h}\right)\frac{3\pi^2}{16}\Delta^2$ and $W(x)$ is the Lambert W function.
The conductance for several different $\Delta$ is plotted in Fig.~\ref{Gcr}.
At the fixed point
$g_{LR}^*$=$1/K$. Thus the decoherence rate (defined in Eq.~(\ref{gamdef}))
for $V\ll T_K$ is,
\begin{eqnarray}
\Gamma = \frac{\pi V}{K}
\end{eqnarray}
and thus apparently small for $K \gg 1$.
However, as we show below, this decoherence rate plays an important role in cutting off the
logarithms in the $T$-matrix.

For $V$=$T$=$0$ the $T$ matrix has a powerlaw cusp, $\rm{Im}
T(\omega)\approx c_1-c_2 |\omega|^\Delta$~\cite{Affleck93,Gan94}.
Eqn.~(\ref{Tpert3}) suggests that this peak will split by $V$ and we also expect it
to be broadened by $\Gamma$. As $\Gamma \ll V$ this splitting will
be observable even for $V \ll T_K$. In the following we will use RG
to calculate this split Kondo resonance for  $V \ll T_K$ (results for
$V\gg T_K$ follow from the literature of the
single-channel Kondo model~\cite{Freqdep}). We assume the symmetrical application
of voltages $\mu_L$=$- \mu_R$=$V/2>0$. For $|\Omega-V/2| \ll V$, and for $V \gg D>|\Omega-V/2|,\Gamma$,
Eq.~(\ref{rgfull}) gives,
\begin{eqnarray}
&&g_{d}(D)
\approx \frac{1}{2}
\left[g^*  - \left(g^{*}-g(V)\right)
\left(\frac{D}{V}\right)^{\Delta/2}
\right]\\
&&\approx \frac{g^*}{2}
\left[1 - \xi  \left(\frac{V}{T_K}\right)^{\Delta} \left(\frac{D}{V}\right)^{\Delta/2}
\right]\label{simply}
\end{eqnarray}
The resummed T-matrix is then given by,
\begin{eqnarray}
&&-\pi \nu \,{\rm Im}\!\!\left[T^{\alpha \alpha^{\prime} \sigma}_R(\Omega)\right]
\approx\frac{3\pi^2}{16}
\left[2 g_{d}^2(max(\Omega-V/2,\Gamma))\right] \nonumber \\
\label{Tsol}
\end{eqnarray}
Expanding Eq.~\ref{Tsol} in powers of the interaction $g$, we have checked that this is
consistent with Eqn.~(\ref{Tpert3}).

For $V<\Omega<T_K$ the $T$-matrix is described by a cusp with
power $\Delta$. Apparently, for $\Gamma< |\Omega-V/2|\ll V$ this
exponent changes to $\Delta/2$ and one may want to interpret this as
a new nonequilibrium scaling regime governed by a $2 K$ channel,
instead of $K$ channel behavior. However, decoherence is so strong
that this regime is `unmeasurable': the extra factor in Eq.~(\ref{simply})
$1>(D/V)^{\Delta/2} \gtrsim (\Gamma/V)^{\Delta/2}\approx
e^{-(\ln K)/K}$ always remains close to
$1$. Moreover, the peak at $\Omega$=$V/2$ is not much higher than the
minimum at $\Omega$=$0$, ${\rm Im}[ T(V/2)-T(0)]/{\rm Im} [T(0)]\sim
\left(\frac{V}{T_K}\right)^\Delta \frac{\ln K}{K}\ll 1 $.  These results are
shown schematically in Fig.~\ref{Gcr} where we have used $max(a,b)$=$\sqrt{a^2+b^2}$.

Above, we have shown that a $2K$ channel fixed point cannot be
stabilized by a finite voltage in an extended regime. However, if
one considers a model where initially  $g_{LR} \ll g_d$ such a
regime becomes accessible. Remarkably, one can even calculate the
nonequilibrium conductance exactly in this regime for
$K$=$1$~\cite{EranSela}. It has also been suggested by one of us~\cite{Rosch01}
that a finite voltage stabilizes the $2K$ channel
regime efficiently for $K$=$1$ for sufficiently small $g_{LR}$. It is therefore
interesting to study this case also for $K\gg 1$.

For $g_{LR}$=$0$ the system is described by a $2 K$
channel fixed point with a Kondo temperature $T_K$=$D_0 (K g_{d0})^K e^{-1/g_{d0}}$,
and voltage has no effect as no currents can
flow. However $g_{LR}$ is a relevant variable for $V$=$0$. While at the scale $T_K$
it is small, $g_{LR}(T_K)\approx \frac{g_{LR0}}{g_{d0}^2 K^2}$, it
grows below $T_K$ with scaling dimension $-1/K$ (compared to the
 exact result \cite{Affleck93}
$-1/(K+1)$), thus inducing a flow back
to the $K$ channel fixed point which is reached at the scale
$T_K^*\approx T_K \left( \frac{g_{LR0}}{K g_{d0}^2} \right)^K $.

New physics can arise in the regime $T_K^*<V<T_K$ governed by the
$2K$-channel fixed point. For example,
the conductance in this regime is given by
$G_m \sim \frac{e^2}{\hbar} g_{LR}(V)^2 \sim   \frac{e^2}{\hbar K^2}\left(
\frac{T_K^*}{V} \right)^{2/K} $
which crosses smoothly over to Eqn.~(\ref{conductance}) for $V\sim
T_K^*$.
Similarly, the decoherence rate in this regime is given by
$\Gamma \sim K V g_{LR}(V)^2 \sim  \frac{V}{K} \left(
\frac{T_K^*}{V} \right)^{2/K}$. To decide whether the range of
validity of the $2 K$ channel regime is enhanced compared to the
equilibrium case we compare $\Gamma$ with $T_K^*$ and $V$ for
$T_K>V>T_K^*$. We find
\begin{eqnarray}
\frac{\Gamma}{T_K^*}\sim \frac{1}{K}
\left(\frac{V}{T_K^*}\right)^{1-2/K},\qquad
 \frac{\Gamma}{V}\sim \frac{1}{K}
 \left(\frac{V}{T_K^*}\right)^{-2/K} \end{eqnarray}
While the ratio $\Gamma/V$ is reduced in this regime, the decoherence
rate is always {\em larger} than $T_K^*$ for $V/T_K^*\gtrsim K$ and
approaches
$\Gamma/V\approx 1/K$ as before for smaller voltages (up to
small corrections of $O(\ln K/K)$). This implies that voltage bias does {\em not}
enhance the regime where $2 K$ channel physics is observable in
contrast to the suggestion of Ref.~\cite{Rosch01}.

{\bf Leads with different voltage differences:} We now discuss the case
where (for  $g_{LR}$=$g_d$) a fraction $p$ and hence $K p$ leads are at voltage $V \ll T_K$, while $K(1-p)$
are at $V=0$. (In~\cite{Potok07} one has $p=1/2$ and $K=2$).
The question arises whether the
decoherence $\Gamma$ again prohibits the flow to new fixed
points for finite $V$. Analyzing as above the logarithmic corrections to
the $T$-matrix in the regime $D<V$ we obtain
\begin{eqnarray}
\frac{dg_0}{d\ln D}&=&-\left[g_0^2 -\frac{K}{2}g_0\{(1-p)g_0^2 + p
  (2g_{dV}^2)\}\right] \nonumber\\
\frac{dg_{dV}}{d\ln D}&=& -\left[g_{dV}^2 -\frac{K}{2}g_{dV}\{(1-p)g_0^2 + p (2g_{dV}^2)\}\right]
\end{eqnarray}
where  $g_{dV}$ and $g_0=2 g_d$  are the coupling constants for the
leads with and without an applied bias voltage respectively.
The initial values $g_{dV}(V)$=$g_0(V)/2$=$g(V)/2$ are obtained from
Eq.~(\ref{rgeq3a}) or~(\ref{gRsol}).
For our initial condition
$g_{dV} < g_0$, the solution apparently flows towards the fixed point
$g_{dV}^*$=$0,g_0^*$=$2/(K(1-p))$ where only the channels without bias
voltage contribute to screening, and powerlaws are governed by the
exponent $\Delta_p$=$2/(K(1-p))$. This is however misleading because the
decoherence stops the flow towards this fixed point for {\em all} $p$.
To see this note that
close to the $K$ channel fixed point $g_{dV}$=$g_0/2$=$1/K$ the RG equations are
$\frac{dg_0}{d\ln D}\approx - 2p/K^2$ and $\frac{dg_{dV}}{d\ln
  D}\approx (1-p)/K^2$. Therefore all changes arising from the flow
from $D=V$ to $D=\Gamma \simeq \pi p V/K$ remain for arbitrary $p$ smaller than
$\frac{2}{K^2} \ln(V/\Gamma) < \frac{2}{K^2} \ln(K^2)\ll 1/K$. Therefore the new
fixed points can never be approached.

In summary we have studied the nonequilibrium overscreened Kondo
problem in the perturbatively solvable limit of a large number of
leads $K\gg 1$.  For this model, current induced decoherence is very
small ($\Gamma/V ={\cal O}(1/K)\ll 1$). However, in this limit the
renormalization group flows also become very slow. Our calculations
show that the net result is that the
flow to any new voltage induced fixed points is stopped very effectively
by the decoherence.  We have made
predictions for the splitting of the T-matrix for $V\ll T_K$ which can
be observed experimentally. An important open question is to study
this nonequilibrium problem in the presence of an external magnetic
field.

{\sl Acknowledgements:}  We thank
E. Sela and  D. Schuricht for useful comments. This work was supported by NSF-DMR (Award No. 0705584 and 1004589)
and by the DFG within SFB 608 and FOR 960.

\end{document}